\documentclass[%
 reprint,
 amsmath,amssymb,
 aps,
]{revtex4-1}

\usepackage{graphicx}% Include figure files
\usepackage{dcolumn}% Align table columns on decimal point
\usepackage{bm}% bold math
\begin{document}

\preprint{APS/123-QED}

\title{$\mathbf{NaCuMoO_4(OH)}$ as a Candidate Frustrated $\bm{J_1}$--$\bm{J_2}$ Chain Quantum Magnet}% Force line breaks with \\

\author{Kazuhiro~Nawa$^1$}
\email{knawa@issp.u-tokyo.ac.jp}
\author{Yoshihiko~Okamoto$^1$}%
\altaffiliation[Present address: ]{Department of Applied Physics, Graduate School of Engineering, Nagoya University, Chikusa, Nagoya 464-8603, Japan}
\author{Akira~Matsuo$^1$}
\author{Koichi~Kindo$^1$}
\author{Yoko~Kitahara$^2$}
\author{Syota~Yoshida$^2$}
\author{Shohei~Ikeda$^2$}
\author{Shigeo~Hara$^3$}
\author{Takahiro~Sakurai$^3$}
\author{Susumu~Okubo$^4$}
\author{Hitoshi~Ohta$^4$}
\author{Zenji~Hiroi$^1$}
\affiliation{%
$^{1}$Institute for Solid State Physics, The University of Tokyo, Kashiwa, Chiba 277-8581, Japan \\
$^{2}$Graduate School of Science, Kobe University, Nada, Kobe 657-8501, Japan \\
$^{3}$Center for Supports to Research and Education Activities, Kobe University, Nada, Kobe 657-8501, Japan \\
$^{4}$Molecular Photoscience Research Center, Kobe University, Nada, Kobe 657-8501, Japan
}%

\date{\today}% It is always \today, today,

\begin{abstract}
In a frustrated $J_1$--$J_2$ chain with the 
nearest-neighbor ferromagnetic interaction $J_1$ and the next-nearest-neighbor antiferromagnetic interaction $J_2$,
novel magnetic states such as a spin-nematic state are theoretically expected.
However, they have been rarely examined in experiments because of the difficulty in obtaining suitable model compounds. 
We show here that the quasi-one-dimensional antiferromagnet $\mathrm{NaCuMoO_4(OH)}$, which comprises edge-sharing $\mathrm{CuO_2}$ chains,
is a good candidate $J_1$--$J_2$ chain antiferromagnet.
The exchange interactions are estimated as $J_1$ =~$-$51~K and $J_2$ =~36~K 
by comparing the magnetic susceptibility, heat capacity, and magnetization data with the data obtained using calculations
by the exact diagonalization method.
High-field magnetization measurements at 1.3~K show a saturation above 26~T 
with little evidence of a spin nematic state expected just below the saturation field,
which is probably due to smearing effects
caused by thermal fluctuations and the polycrystalline nature of the sample.
%\begin{description}
%\end{description}
\end{abstract}

\pacs{Valid PACS appear here}% PACS, the Physics and Astronomy
\maketitle

%\tableofcontents

Low-dimensional quantum spin systems with geometrical frustration and/or competing magnetic interactions have attracted much attention in the field of magnetism.
Low dimensionality, quantum fluctuations, and frustration are three ingredients that may effectively suppress conventional
magnetic order
and lead us to unconventional magnetic order or exotic ground states such as a quantum spin liquid\cite{1DQSL, QSL}.

A frustrated $J_1$--$J_2$ chain of spin 1/2 defined as
\begin{equation}
\label{J1J2}
\mathcal{H} = J_1 \sum _{l} \mathbf{s}_l \cdot \mathbf{s}_{l+1} +
J_2 \sum _{l} \mathbf{s}_l \cdot \mathbf{s}_{l+2} - h \sum _{l} s_{l}^{z}
\end{equation}
provides us with  an interesting example:
the competition between the nearest-neighbor (NN) ferromagnetic interaction $J_1$
and the next-nearest-neighbor (NNN) antiferromagnetic interaction $J_2$
causes various quantum states in magnetic fields $h$\cite{1Dtheory00, 1Dtheory0, 1Dtheory1, 1Dtheory2, 1Dtheory3}.
Realized in low fields is a long-range order of vector chirality defined as
$(\mathbf{s}_l \times \mathbf{s}_{l+n})_z \ (n =~1, 2)$.
As the field increases, 
spin correlations change markedly because bound magnon pairs are stabilized by ferromagnetic $J_1$.
The bound magnon pairs form a spin density wave (SDW)
in medium fields, whereas,
in high fields just below the saturation of magnetization,
they exhibit Bose--Einstein condensation into quantum multipolar states\cite{nematic1, nematic2, nematic3, nematic4}.
One of the multipolar states expected just below the saturation is a quadrupolar state of magnon pairs called a spin nematic state, 
analogous to nematic liquid crystals.

To explore these quantum states theoretically predicted for the frustrated $J_1$--$J_2$ chain, 
many experimental studies have been performed on quasi-1D compounds such as
$\mathrm{Li_2ZrCuO_4}$\cite{Li2ZrCuO4_0, Li2ZrCuO4},
$\mathrm{Rb_2Cu_2Mo_3O_{12}}$\cite{Rb2Cu2Mo3O12_0, Rb2Cu2Mo3O12},
$\mathrm{PbCu(SO_4)(OH)_2}$\cite{PbCuSO4OH_0, PbCuSO4OH_2, PbCuSO4OH_3},
$\mathrm{LiCuSbO_4}$\cite{LiCuSbO4}, 
$\mathrm{LiCu_2O_2}$\cite{LiCu2O2, LiCu2O2_neu, LiCu2O2_3}, and
$\mathrm{LiCuVO_4}$\cite{magnetization, HFNMR, LiCuVO4, cryst, neutron0, NMR1, neutron1, neutron2, NMR2},
the key parameters of which are listed in Table~\ref{tab:param}.
These compounds commonly have edge-sharing $\mathrm{CuO_2}$ chains made of $\mathrm{CuO_6}$ octahedra.
NN Cu spins are magnetically coupled with each other through two superexchange Cu--O--Cu paths with approximately 90$^\circ$ bond angles,
while NNN Cu spins are coupled through two super-superexchange Cu--O--O--Cu paths.
Thus, according to the Goodenough--Kanamori rule, $J_1$ should be ferromagnetic while $J_2$ can be antiferromagnetic.
This is in fact the case for these candidate compounds, which causes frustration in the $J_1$--$J_2$ chains.

\begin{table}[t]
\caption{
\label{tab:param} Candidate compounds for the $J_1$--$J_2$ chain system.
Listed are the nearest-neighbor intrachain interaction $J_1$, the next-nearest-neighbor interaction $J_2$,
the bond angles of Cu-O-Cu paths for $J_1$, the antiferromagnetic transition temperature at zero field $T_\mathrm{N}$,
and the saturation field $H_\mathrm{s}$.}
\label{coeff}
\begin{center}
\begin{tabular}
{cccccccc}
Compound & $J_1$, $J_2$ & $\angle$ Cu-O-Cu & $T_\mathrm{N}$ & $H_\mathrm{s}$ \\ 
 & (K) & (deg) & (K) & (T)  \\ \hline 
$\mathrm{Li_2ZrCuO_4}$\cite{Li2ZrCuO4_0, Li2ZrCuO4}
& $-$151, 35 & 94.1 & 6.4 & - \\
$\mathrm{Rb_2Cu_2Mo_3O_{12}}$\cite{Rb2Cu2Mo3O12_0, Rb2Cu2Mo3O12}
& $-$138, 51 & 89.9, 101.8 & $<$ 2 & 14 \\
& & 91.9, 101.1 &  & \\
$\mathrm{PbCuSO_4(OH)_2}$\cite{PbCuSO4OH_0, PbCuSO4OH_2, PbCuSO4OH_3} & $-$100, 36
& 91.2, 94.3 & 2.8 & 5.4 \\
$\mathrm{LiCuSbO_4}$\cite{LiCuSbO4} & $-$75, 34 & 89.8, 95.0 &$<$ 0.1 & 12 \\
& & 92.0, 96.8 & & & \\
$\mathrm{LiCu_2O_2}$\cite{LiCu2O2, LiCu2O2_neu, LiCu2O2_3} & $-$69, 43 & 92.2, 92.5 & 22.3 & 110 \\ %($H \parallel c$)
$\mathrm{LiCuVO_4}$\cite{magnetization, HFNMR, LiCuVO4, cryst, neutron0, NMR1, neutron1, neutron2, NMR2} & $-$19, 44 & 95.0 & 2.1 & 44.4 \\ \hline %($H \parallel c$)
$\mathrm{NaCuMoO_4(OH)}$ & $-$51, 36 & 92.0, 103.6 & 0.59 & 26 \\ \hline
\end{tabular}
\end{center}
\end{table}
Among these compounds,
the most often studied is $\mathrm{LiCuVO_4}$ with $J_1$ =~$-$19~K and $J_2$ =~44~K\cite{neutron0}.
It has been shown using large single crystals that 
$\mathrm{LiCuVO_4}$ exhibits an incommensurate helical order at low fields\cite{neutron0, NMR1, neutron1, neutron2, NMR2},
which may be a 3D analogue of the vector chirality order in the $J_1$--$J_2$ chain,
and a longitudinal SDW order at intermediate fields\cite{NMR1, neutron1, neutron2, NMR2}.
Furthermore, a spin nematic phase, a 3D analogue of the spin nematic state, has been suggested slightly below the saturation field of 44.4~T at $H \parallel c$,
where the magnetization shows a linear field dependence\cite{magnetization}.
However, the presence of the spin nematic phase is still unclear because of the high saturation field.
Only NMR experiments were performed around the saturation, which revealed that the majority of magnetic moments were already saturated above 41.4~T,
where the spin nematic phase was suggested from magnetization measurements. 
This discrepancy is likely due to crystal defects such as Li deficiency\cite{HFNMR}.
The other candidate compounds thus far studied also have some problems, such as disorder effects and the lack of large single crystals.
Thus, an alternative compound is required for further experimental study of the $J_1$--$J_2$ chain.

\begin{figure}[t]
\centering
\includegraphics[width=7.1cm]{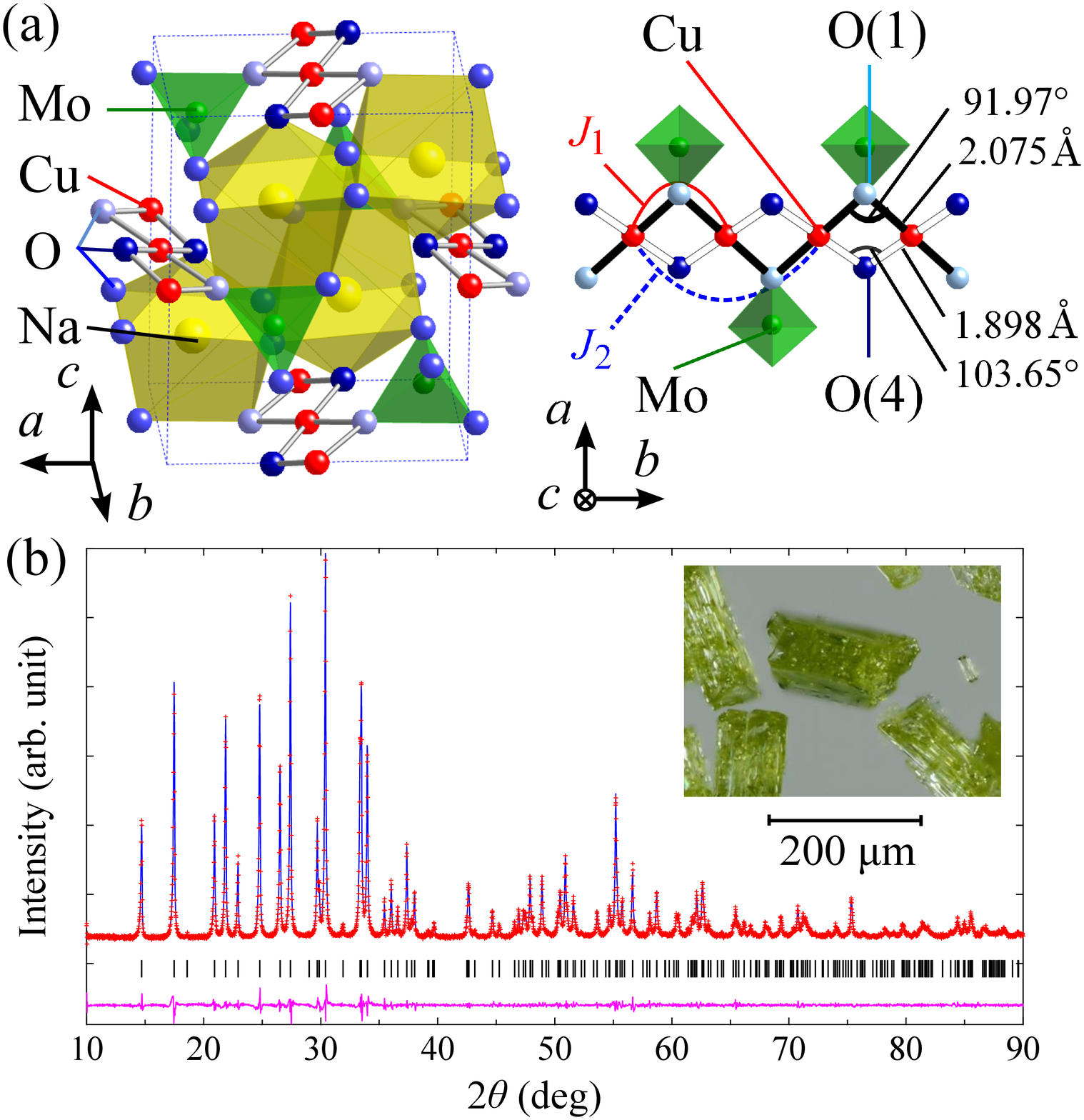}
\caption{
\label{fig:struct} (Color online) (a) Crystal structure of $\mathrm{NaCuMoO_4(OH)}$ (left) and a local environment around a $\mathrm{CuO_2}$ chain (right)
based on the structural parameters reported in Ref. 32.
(b) Observed (red cross) and calculated (blue curve) powder XRD patterns (Cu K$_{\alpha_1}$ radiation)
of a polycrystalline sample of $\mathrm{NaCuMoO_4(OH)}$.
Additional contributions from Cu $K_{\alpha_2}$ radiation have been analytically removed.
The positions of reflections and the difference in intensity
between the observed and calculated patterns are indicated by vertical black lines  and a magenta curve at the bottom, respectively.
The inset shows a photograph of a small single crystal of $\mathrm{NaCuMoO_4(OH)}$.}
\end{figure}

Here, we show that $\mathrm{NaCuMoO_4(OH)}$ is a good candidate compound that meets various experimental requirements.
$\mathrm{NaCuMoO_4(OH)}$ was first prepared hydrothermally by Moini et al. in 1986 \cite{NaCuMoO4OH}.
It crystallizes in an orthorhombic structure with the space group $Pnma$,
which is isomorphous with that of the natural mineral Descloizite $\mathrm{PbZnVO_4(OH)}$\cite{PbZnVO4OH}.
As shown in Fig.~\ref{fig:struct}(a), there is a $\mathrm{CuO_2}$ chain that may represent a $J_1$--$J_2$ chain,
similar to that observed in related compounds.
We discover that $\mathrm{NaCuMoO_4(OH)}$ is a quasi-1D frustrated antiferromagnet
with $J_1$ =~$-$51~K, $J_2$ =~36~K, and $T_\mathrm{N}$ =~0.59~K. 
In addition, we show that the reasonably low saturation field of 26~T of this compound makes it promising
for investigating an exotic spin nematic phase expected in the $J_1$--$J_2$ chain system.

A polycrystalline sample of $\mathrm{NaCuMoO_4(OH)}$ was synthesized by the hydrothermal method.
First, 3.807~g of 5~M NaOH aqueous solution (16.0~mmol of NaOH) was diluted by adding water to a volume of 10 ml.
Then, 1.155~g of $\mathrm{MoO_3}$ (8.0~mmol) and 0.8315~g of $\mathrm{CuSO_4}$ $\cdot$ $5\mathrm{H_2O}$ (3.3 mmol) were added.
The mixed solution was put in a Teflon beaker of 30 ml volume, placed in a stainless steel autoclave, and heated at 240~$^\circ$C for 48 h.
An aggregate of small yellowish green crystals 
having a rodlike shape and a typical size of 0.1$\times$0.1$\times$0.2~$\mathrm{mm}^3$ [inset of Fig. 1(b)] was obtained.
The crystals were filtered, washed with water and ethanol, and dried at room temperature.
To estimate the lattice contribution in heat capacity,
a nonmagnetic analogue $\mathrm{NaZnMoO_4(OH)}$ was also prepared in a similar way.

Sample characterization was performed by powder X-ray diffraction (XRD) analysis using Cu $K_\alpha$ radiation (RINT-2000, Rigaku),
by chemical analysis using inductively coupled plasma spectrometry (JY138KH, Horiba),
and by thermal gravimetry (TG-DTA2020SAH, Bruker AXS).
A powder XRD pattern from crashed crystals is shown in Fig.~\ref{fig:struct}(b).
In a whole powder pattern fitting using a program PDXL (Rigaku),
all the peaks are indexed to reflections allowed for the space group $Pnma$
with the lattice constants $a$ =~7.7338(3)~\AA, $b$ =~5.9678(2)~\AA, and $c$ =~9.5091(3)~\AA,
which are close to those previously reported: $a$ =~7.726(2)~\AA, $b$ =~5.968(2)~\AA, and $c$ =~9.495(3)~\AA\cite{NaCuMoO4OH}.
The chemical compositions of Na, Mo, and Cu are 8.3(1), 23.6(1), and 35.3(2)~wt$\%$,
respectively, which are close to the stoichiometric compositions of 8.7, 24.1, and 36.4~wt$\%$;
the small deviation may be due to the inclusion of small amounts of byproducts.
A dehydration reaction with a weight loss of 3.3(1)\% was observed above 350 C$^\circ$,
which means that nearly half mol of $\mathrm{H_2O}$ has been lost as expected from the chemical composition.
Thus, we have successfully obtained $\mathrm{NaCuMoO_4(OH)}$
for detailed characterizations of its magnetic properties.

Magnetic susceptibility was measured in a SQUID magnetometer (MPMS, Quantum Design),
and magnetization was measured up to 50 T in a pulse magnet
at the Ultra High Magnetic Field Laboratory of the Institute for Solid State Physics at the University of Tokyo\cite{Highfieldmagnetization}.
Heat capacity was measured by the relaxation method (PPMS, Quantum Design).
The $g$ factor of the paramagnetic state was estimated by multifrequency high-field ESR measurements
up to 520 GHz at Kobe University\cite{HFEPR}
instead of by conventional X-band ESR measurements.
Heat capacity measurements were performed on a thin pellet of a powdered sample,
and all the other measurements were performed on an aggregate of single crystals.

\begin{figure}[t]
\centering
\includegraphics[width=9cm]{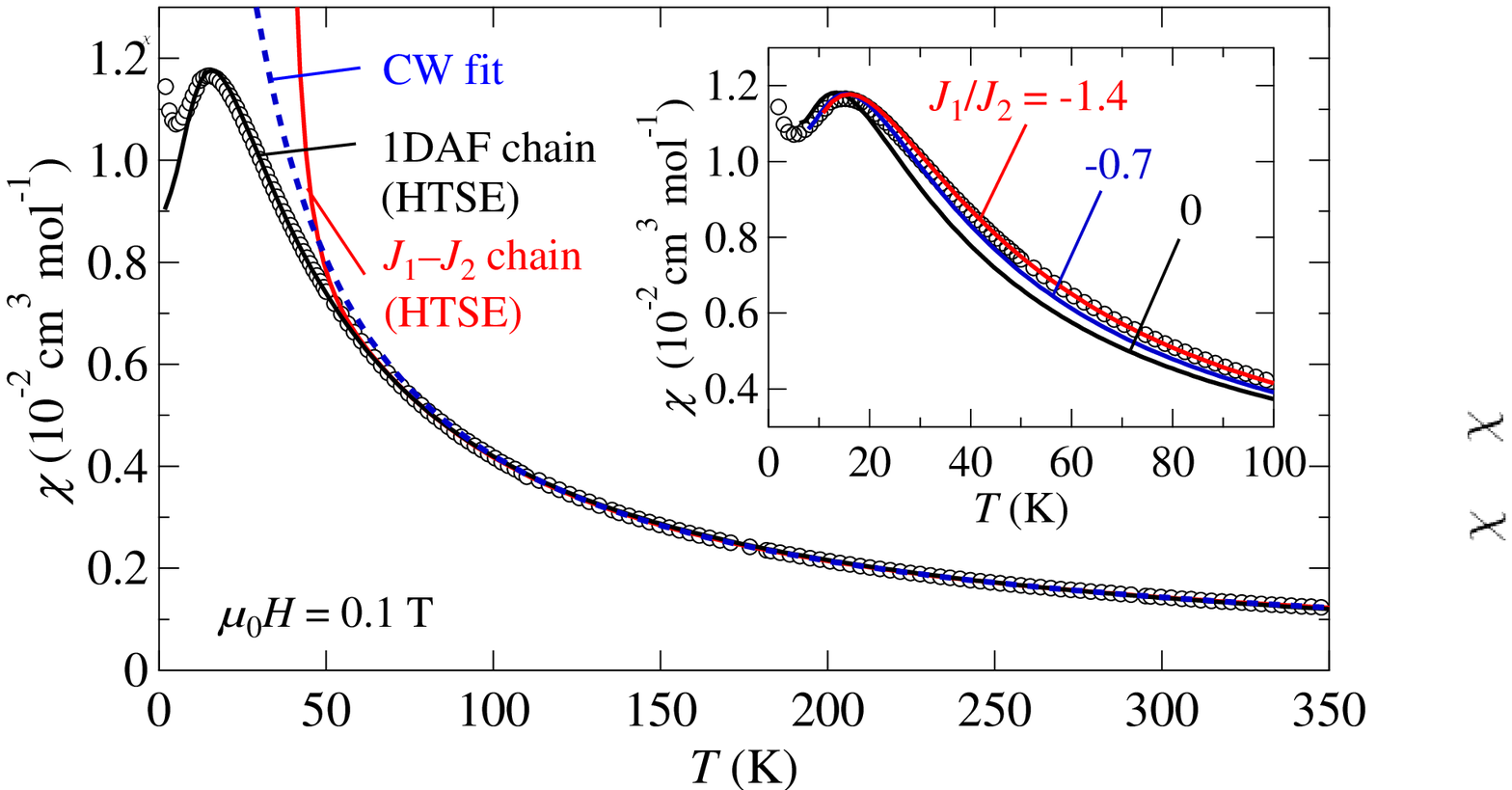}
\caption{\label{fig:chi} (Color online) Temperature dependences of magnetic susceptibility.
The blue dashed cuves represent a Curie--Weiss (CW) fit, and 
the black and red solid curves represent fits to calculations based on the high-temperature series expansion (HTSE) for uniform 1D antiferromagnetic (1DAF)\cite{Johnston}
and frustrated $J_1$--$J_2$ chains\cite{HTSE}, respectively.
The inset shows the magnetic susceptibility compared with these obtained using calculations by the exact diagonalization method for $J_1$--$J_2$ chains
with $J_1/J_2$ =~$-$1.4 (red curve), $-$0.7 (blue curve), and 0 (black curve),
where the $g$ factor and temperature-independent contribution are fixed to 
$g$ =~2.11 determined from the ESR measurement and $\chi_0$ =~1.2$\times$10$^{-5}$~cm$^3$~mol$^{-1}$ from the HTSE fit, respectively.
}
\end{figure}

The temperature dependence of magnetic susceptibility is shown in Fig.~\ref{fig:chi}.
No anomaly indicative of a long-range order is observed above 2~K,
while a broad peak is observed at 14~K, indicating the presence of a 1D antiferromagnetic correlation. 
The magnetic susceptibility in the range of 150--350 K is fitted to the sum of a Curie--Weiss contribution and a temperature-independent contribution $\chi_0$,
\begin{equation}
\chi(T) = \frac{N g^2  \mu^2_\mathrm{B} S(S+1)}{3 k_\mathrm{B} (T - \theta)} + \chi_\mathrm{0} \label{CW},
\end{equation}
where $g$ is the Lande $g$ factor, $\mu_\mathrm{B}$ is the Bohr magnetron, $k_\mathrm{B}$ is the Boltzmann constant, and $\theta$ is the Weiss temperature.
The fitting shown by the dotted line in Fig.~\ref{fig:chi} yields
$\theta$ =~$-$5.0(5)~K, $g =$~2.18(1), and $\chi_\mathrm{0}$ =~$-$2.5(4)$\times$10$^{-5}$~cm$^3$~mol$^{-1}$.

$\chi$ in the range of 8--350~K is alternatively fitted to a 1D antiferromagnetic (1DAF) chain model\cite{Johnston}.
The calculated curve well reproduces $\chi$, particularly the broad peak at 14~K,
and yields $J_\mathrm{1DAF} =$~24.4(1)~K, $g =$~2.30(1), and $\chi_\mathrm{0}$ =~$-$1.67(1)$\times$10$^{-4}$~cm$^3$~mol$^{-1}$.
However, this fitting suffers from the following two inconsistencies.
First, $g$ of 2.30 is too large for the powder average for Cu$^{2+}$ ions, typically 2.1--2.2,
and $\chi_\mathrm{0}$ is too small compared with the diamagnetic susceptibility from core electrons,
$\chi_\mathrm{dia}$ =~$-$8.3$\times$10$^{-5}$~cm$^3$~mol$^{-1}$;
$\chi_\mathrm{0}$ must be larger than $\chi_\mathrm{dia}$
since there must be an additional positive contribution from the Van Vleck paramagnetism.
Second, the Weiss temperature expected in the mean field theory is $\theta$ =~$-J_\mathrm{1DAF}/2$ =~$-$12~K in the 1DAF chain, 
which is significantly different from $\theta$ =~$-$5.0~K  from the Curie--Weiss fit.
Similar discrepancies have been observed in $\mathrm{LiCuVO_4}$:
a fit to the 1DAF chain model gives a larger $g$ than that determined by ESR measurements\cite{ueda} and
a smaller $-J_\mathrm{1DAF}/2$ than $\theta$ for the Curie--Weiss fit\cite{neutron0,ueda}.
Thus, there must be additional ferromagnetic couplings in these compounds.

Provided that there are two magnetic interactions,
ferromagnetic $J_1$ and antiferromagnetic $J_2$, in $\mathrm{NaCuMoO_4(OH)}$,
we have determined $J_1$ and $J_2$ by analyzing $\chi$ more elaborately on the bases of simulations by 
the high-temperature series expansion (HTSE)\cite{HTSE} and exact diagonalization (ED) method.
First, we have determined the $g$ factor by ESR experiments.
Absorption lines observed at 173 K at frequencies between 200 and 520 GHz
were well reproduced by single Lorentzian curves with linewidths of about 2 T.
$g$ is estimated to be 2.11(2) from a linear relation between the resonant frequency and the field. 

Fitting to the $J_1$--$J_2$ chain model based on the HTSE in the range of 100--350~K using $g$ =~2.11 yields
$J_\mathrm{1}$ =~-61(2)~K, $J_\mathrm{2}$ =~41(2)~K, and $\chi_\mathrm{0}$ =~1.2(2)$\times$10$^{-5}$~cm$^3$~mol$^{-1}$.
This estimation for $J_1$ and $J_2$, however, may not be reliable, because the HTSE is applicable only at high temperatures, while $\chi_0$ must be reliable.
In contrast, ED calculations can simulate $\chi$ down to lower temperatures: 
our full diagonalization for $N$ =~18 spins in the periodic boundary condition using the ALPS package\cite{ALPS} may be reliable down to $T \sim 0.4 J_2$.
Assuming $\chi_0$ =~1.2$\times$10$^{-5}$~cm$^3$~mol$^{-1}$ from the HTSE fit,
we have obtained a best fit to the experimental data at $J_1$ =~$-$51~K and $J_2$ =~36~K ($J_1$/$J_2$ =~$-$1.4).
$\chi$ values compared with a series of calculations for $J_1$/$J_2$ =~$-$1.4, $-$0.7, and 0 are shown in the inset of Fig.~\ref{fig:chi}.
We can determine the $J_1$ and $J_2$ almost uniquely to reproduce the whole $\chi$.

\begin{figure}[t]
\centering
\includegraphics[width=8.5cm]{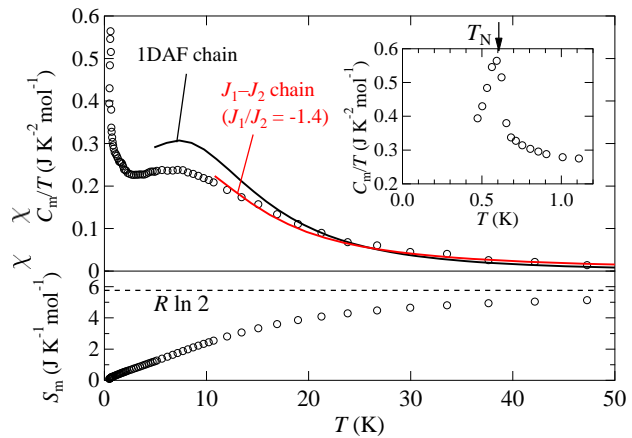}
\caption{\label{fig:HC} 
(Color online) Temperature dependences of magnetic heat capacity divided by temperature and magnetic entropy.
The black and red curves show heat capacities for a uniform 1DAF chain with $J$ =~24.4 K
and a frustrated $J_1$--$J_2$ chain with $J_1$ =~$-$51~K and $J_2$ =~36~K \ ($J_1/J_2 =~-1.4$)
calculated by the exact diagonalization method, respectively.
The inset expands the low-temperature part, where a sharp anomaly indicates a phase transition.}
\end{figure}

The heat capacity $C$ is another important thermodynamic quantity that carries information about magnetic properties.
The lattice contribution of $\mathrm{NaCuMoO_4(OH)}$ has been estimated from the heat capacity of $\mathrm{NaZnMoO_4(OH)}$.
Taking into account a  possible difference in the Debye temperature between the two compounds,
the $C$--$T$ curve of $\mathrm{NaZnMoO_4(OH)}$ has been expanded along the $T$ axis by a factor of 1.07
so that the high-temperature parts above 150~K coincide between them.
The magnetic contribution $C_m$ is obtained by subtracting the expanded curve.
$C_m$ divided by temperature, $C_m/T$, which is shown in Fig.~\ref{fig:HC}, exhibits a broad peak at 8~K, indicating the development of a short-range magnetic order,
and then a sharp increase followed by a cusp at 0.59~K, which gives clear evidence of a long-range order.
$C_m/T$ down to low temperatures
is not reproduced by ED calculations for a 1DAF chain with $J$ =~24.4~K
but for a $J_1$--$J_2$ chain with the same $J_1$ and $J_2$ used in the $\chi$ fitting,
which clearly demonstrates the reliability of our estimation.
The transition temperature $T_\mathrm{N}$ of 0.59~K is as low as about 1$\%$ of $J_2$, indicating a good one-dimensionality in magnetic interactions.
Note that the one-dimensionality is better in the present compound than in $\mathrm{LiCuVO_4}$:
the $T_\mathrm{N}$ of $\mathrm{LiCuVO_4}$ is 2.1~K, which corresponds to about 5$\%$ of $J_2$.
Assuming that $C_m/T$ decreases to 0 linearly below 0.5~K as expected from the high-temperature curve, 
the magnetic entropy $S_m$ below $T_\mathrm{N}$ is estimated to be 0.15~J mol$^{-1}$, which is only 2.6$\%$ of the total entropy of $R \ln 2$ for spin 1/2.
This confirms the good one-dimensionality of the present compound.

\begin{figure}[t]
\centering
\includegraphics[width=8.5cm]{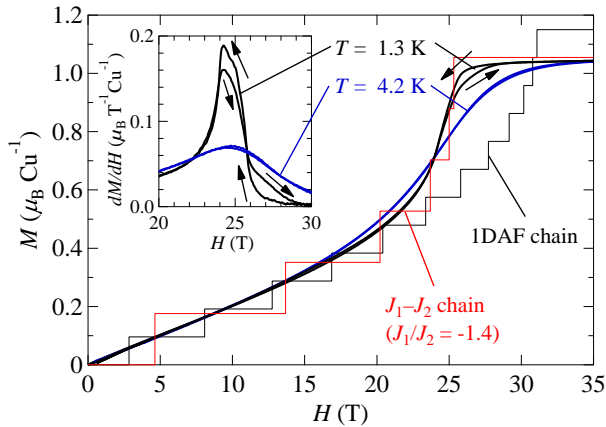}
\caption{\label{fig:MH} (Color online) Magnetization curves recorded at 1.3~K (black curve) and 4.2~K (blue curve)
upon increasing and then decreasing magnetic field in short magnetic
pulses of a few milliseconds.
An aggregate of small crystals with random orientation was used for each measurement.
The stepwise black line shows a calculated magnetization curve for a uniform 1DAF chain with $J =~24.4$~K and $g =~2.30$,
and the red one is for a frustrated $J_1$--$J_2$ chain with $J_1$ =~$-$51~K, $J_2$ =~36~K, and $g =~2.11$.
Both calculations were performed by the Lanczos method for $N$ =~24 sites.
The inset shows the corresponding field-derivative curves near the saturation.}
\end{figure}

Magnetization measurements up to 50~T were performed to search for a spin nematic phase.
Magnetization curves measured at 1.3 and 4.2~K are shown in Fig.~\ref{fig:MH},
which are compared with those calculated by the Lanczos method for $N$ =~24 sites using the ALPS package\cite{ALPS}.
The small hysteresis in the magnetization curve may be due to a magnetocaloric effect under a quasi-adiabatic condition.
The magnetization curve at 1.3~K rapidly increases at 23~T and almost saturates above 26~T,
while the 4.2~K curve rises more gradually owing to thermal fluctuations. 
The saturation moment of 1.05~$\mu_B$/Cu is consistent with the $g$ factor of 2.11(2).
The calculated curve for the $J_1$--$J_2$ chain model with the same parameters from the analyses of $\chi$
reproduces the 1.3~K curve very well, while that for the 1DAF chain does not.
Note that the calculations assume $T$ =~0,
so that a thermal smearing effect should always be taken into account when compared with experiments.
The saturation field $H_\mathrm{s}$ is calculated to be 25.4~T using the equation
$H_\mathrm{s}$ =~$\{ J_1 + 3J_2 - J_2^2/(J_1 - J_2) \}
\ k_\mathrm{B}/(2 g \mu_B)$ \cite{1Dtheory00,1Dtheory0},
which is in good agreement with the experimental one.

In a single crystal of $\mathrm{LiCuVO_4}$,
a linear variation in magnetization has been observed just below the saturation at 1.3~K,
which may be associated with a spin nematic order\cite{magnetization}.
Such a linear variation is not discernible in Fig.~\ref{fig:MH} for the present compound.
The field derivatives of the magnetization curves at 1.3 K show asymmetric peaks at 24--25~T.
This is partly because we used an aggregate of small crystals:
an anisotropy in the $g$ factor should cause a distribution in $H_\mathrm{s}$
so that a linear variation below $H_\mathrm{s}$ could be averaged to disappear.
In addition, the measurement temperature of 1.3~K may not be low enough compared with $T_\mathrm{N}$ =~0.59~K
to stabilize a spin nematic phase.
Note that the temperature of 1.3 K is lower than $T_\mathrm{N}$ =~2.1~K for $\mathrm{LiCuVO_4}$.
Further experiments using a large single crystal at temperatures as low as $T_\mathrm{N}$
are necessary to obtain evidence of the spin nematic phase in $\mathrm{NaCuMoO_4(OH)}$.

We have shown that $\mathrm{NaCuMoO_4(OH)}$ with $J_1$ =~$-$51~K and $J_2$ =~36~K can be a good model compound for the $J_1$--$J_2$ chain system.
Let us compare $\mathrm{NaCuMoO_4(OH)}$ with the other model compounds listed in Table~\ref{tab:param}.
It is found in these compounds that $J_2$ does not change so much at 30--50 K, while $-J_1$ varies largely
at 20--150~K.
This is because $J_2$ occurs by the super-super exchange interaction via the Cu--O--O--Cu path,
while $J_1$ by the superexchange interaction via the Cu--O--Cu path, only the latter of which is sensitive to local structures;
$J_1$ becomes ferromagnetic when the Cu--O--Cu angle is close to 90$^\circ$
and changes into antiferromagnetic when the Cu--O--Cu angle exceeds 95--98$^\circ$\cite{ang, KG}.
For instance, a bond angle close to 90$^\circ$ leads to a large ferromagnetic $J_1$ of $-$138 K in $\mathrm{Rb_2Cu_2Mo_3O_{12}}$, 
while the 95.0$^\circ$ bond angle of $\mathrm{LiCuVO_4}$ gives $J_1$ =~$-$19~K.
In $\mathrm{NaCuMoO_4(OH)}$, two types of Cu--O--Cu path are present,
one passing through O(1) with a 92.0$^\circ$ bond angle and the other passing through O(4) 
connected to hydrogen with a 103.7$^\circ$ bond angle (Fig.~\ref{fig:struct}).
The moderately large ferromagnetic $J_1$ =~$-$51~K may be attained from the dominant contribution of
the Cu--O(1)--Cu path.

The good combination of $J_1$ and $J_2$ in $\mathrm{NaCuMoO_4(OH)}$
provides us with a better opportunity for studying the physics of the $J_1$--$J_2$ chain.
$H_\mathrm{s}$ =~26~T in $\mathrm{NaCuMoO_4(OH)}$ is much smaller
than $H_\mathrm{s}$ =~44.4~T in $\mathrm{LiCuVO_4}$,
and is accessible in various experiments such as magnetization, NMR, and even neutron scattering experiments.
One more important requirement for a good candidate compound is the availability of a large and clean single crystal.
Among the compounds shown in Table~\ref{tab:param}, large single crystals have been obtained only for
$\mathrm{PbCuSO_4(OH)_2}$\cite{PbCuSO4OH_0,  PbCuSO4OH_2, PbCuSO4OH_3},
$\mathrm{LiCu_2O_2}$\cite{LiCu2O2, LiCu2O2_neu, LiCu2O2_3}, and
$\mathrm{LiCuVO_4}$\cite{magnetization, HFNMR, LiCuVO4, cryst, neutron0, NMR1, neutron1, neutron2, NMR2}.
However, "a cleanness" of these crystals seems unsatisfactory: a natural crystal of $\mathrm{PbCuSO_4(OH)_2}$ is contaminated by impurities,
and the two Li-containing crystals seem to suffer from Li deficiency or an interchange between Li and Cu atoms\cite{cryst}.
In contrast, $\mathrm{NaCuMoO_4(OH)}$ shows no such problems as the Na ion is much less mobile in crystals than the Li ion,
and its large ionic radius prevents intersite mixing with Cu.
Therefore, $\mathrm{NaCuMoO_4(OH)}$ can be an ideal compound for the $J_1$--$J_2$ chain quantum magnet in various aspects.
We continue our effort in obtaining a larger single crystal by tuning growth conditions.
In the future, we will clarify the physics of the $J_1$--$J_2$ chain, particularly the
nature of the spin nematic phase by $^{23}$Na NMR experiments and others on sizable single crystals
of $\mathrm{NaCuMoO_4(OH)}$.

In summary, we have investigated the magnetic susceptibility, heat capacity, and magnetization of the quasi-1D quantum antiferromagnet $\mathrm{NaCuMoO_4(OH)}$.
By comparing them with those obtained using calculations by the exact diagonalization method,
it is shown that $\mathrm{NaCuMoO_4(OH)}$ is a good candidate frustrated $J_1$--$J_2$ magnet:
$J_1$ =~$-$51~K, $J_2$ =~36~K, $T_N$ =~0.59~K, and $H_\mathrm{s}$ =~26~T
 (much smaller than 44.4~T for $\mathrm{LiCuVO_4}$).
Although our magnetization measurements at 1.3~K using an aggregate of small crystals
have failed to obtain evidence of the spin nematic order,
we think that our future experiments at lower temperatures using a large single crystal would uncover the
intriguing physics of the frustrated $J_1$--$J_2$ chain.

\begin{acknowledgments}
We thank M. Koike and M. Isobe for chemical analyses and M. Takigawa, G. J. Nilsen, and H. Ishikawa for fruitful discussions.
\end{acknowledgments}


\begin{thebibliography}{99}
\bibitem{1DQSL} H.-J. Mikeska and A. K. Kolezhuk, in \textit{Quantum Magnetism},
ed. U. Schollw\"ock et al., Lecture Notes in Physics Vol. 645 (Springer-Verlag, Berlin, 2004) p. 1
\bibitem{QSL} L. Balents, Nature \textbf{464}, 199 (2010).
\bibitem{1Dtheory00} A. V. Chubukov, Phys. Rev. B \textbf{44}, 4693 (1991).
\bibitem{1Dtheory0} L. Kecke, T. Momoi, and A. Furusaki, Phys. Rev. B \textbf{76}, 060407 (2007).
\bibitem{1Dtheory1} T. Vekua, A. Honecker, H.-J. Mikeska, and F. Heidrich-Meisner, Phys. Rev. B \textbf{76}, 174420 (2007).
\bibitem{1Dtheory2} T. Hikihara, L Kecke, T. Momoi, and A. Furusaki, Phys. Rev. B \textbf{78}, 144404 (2008).
\bibitem{1Dtheory3} J. Sudan, A. L\^uscher, and A. M. L\^auchli, Phys. Rev. B \textbf{80}, 140402 (2009).
\bibitem{nematic1} M. E. Zhitomirsky and H. Tsunetsugu, Europhys. Lett. \textbf{92}, 37001 (2010).
\bibitem{nematic2} M. Sato, T. Hikihara, and T. Momoi, Phys. Rev. Lett. \textbf{110}, 077206 (2013).
\bibitem{nematic3} O. A. Starykh and L. Balents, Phys. Rev. B \textbf{89}, 104407 (2014).
\bibitem{nematic4} H. T. Ueda and K. Totsuka, cond-mat arXiv, 1406.1960v1. 
\bibitem{Li2ZrCuO4_0} C. Dussarrat, G. C. Mather, V. Caignaert, B. Domen\`es, J. G. Fletcher, and A. R. West, J. Solid. State Chem. \textbf{166}, 311 (2002).
\bibitem{Li2ZrCuO4} S.-L. Drechsler, O. Volkova, A. N. Vasiliev, N. Tristan, J. Richter, M. Schmitt, H. Rosner, J. M\'alek, R. Klingeler, A. A. Zvyagin,
and B. Bu\"chner, Phys. Rev. Lett. \textbf{98}, 077202 (2007).
\bibitem{Rb2Cu2Mo3O12_0} S. F. Solodovnikov and Z. A. Solodovnikova,  J. Struct. Chem. \textbf{38}, 765 (1997).
\bibitem{Rb2Cu2Mo3O12} M. Hase, H. Kuroe, K. Ozawa, O. Suzuki, H. Kitazawa, G. Kido, and T. Sekine, Phys. Rev. B \textbf{70}, 104426 (2004).
\bibitem{PbCuSO4OH_0} H. Effenberger, Mineral. Petrol. \textbf{36}, 3 (1987).
\bibitem{PbCuSO4OH_2} A. U. B. Wolter, F. Lipps, M. Schapers, S.-L. Drechsler, S. Nishimoto, R. Vogel, V. Kataev, B. Buchner, H. Rosner,
M. Schmitt, M. Uhlarz, Y. Skourski, J. Wosnitza, S. Sullow, and K. C. Rule, Phys. Rev. B \textbf{85}, 014407 (2012).
\bibitem{PbCuSO4OH_3} B. Willenberg, M. Sch\"{a}pers, K. C. Rule, S. S\"{u}llow, M. Reehuis, H. Ryll, B. Klemke, K. Kiefer, W. Schottenhamel, B. B\"{u}chner, B. Ouladdiaf, M. Uhlarz, R. Beyer, J. Wosnitza, and A. U. B. Wolter, Phys. Rev. Lett. \textbf{108}, 117202 (2012).
\bibitem{LiCuSbO4} S. E. Dutton, M. Kumar, M. Mourigal, Z. G. Soos, J.-J. Wen, C. L. Broholm, N. H. Andersen, Q. Huang, M. Zbiri, R. Toft-Petersen, and R. J. Cava, Phys. Rev. Lett. \textbf{108}, 187206 (2012).
\bibitem{LiCu2O2} R. Berger, A. Meetsma, and S. van Smaalen, J. Less-Common. Met. \textbf{175}, 119 (1991).
\bibitem{LiCu2O2_neu} T. Masuda, A. Zheludev, B. Roessli, A. Bush, M. Markina, and A. Vasiliev, Phys. Rev. B \textbf{72}, 014405 (2005).
\bibitem{LiCu2O2_3} A. A. Bush, V. N. Glazkov, M. Hagiwara, T. Kashiwagi, S. Kimura, K. Omura, L. A. Prozorova, L. E. Svistov, A. M. Vasiliev, and A. Zheludev, Phys. Rev. B \textbf{85}, 054421 (2012).
\bibitem{LiCuVO4} M. A. Lafontaine, M. Leblanc, and G. Ferey, Acta. Cryst. \textbf{C45}, 1205 (1989).
\bibitem{cryst} A. V. Prokofiev, I. G. Vasilyeva, V. N. Ikorskii, V. V. Malakhov, I. P. Asanov, and W. Assmus,
J. Solid State Chem. \textbf{177}, 3131 (2004).
\bibitem{neutron0} M. Enderle, C. Mukherjee, B. F\aa k, R. K. Kremer, J.-M. Broto, H. Rosner, S.-L. Drechsler, J. Richter, J. Malek, A. Prokofiev,
W. Assmus, S. Pujol, J.-L. Raggazzoni, H. Rakoto, M. Rheinst\^adter, and H. M. R\o nnow, Europhys. Lett. \textbf{70}, 237 (2005).
\bibitem{NMR1} N. B\"{u}ttgen, H. -A. Krug von Nidda, L. E. Stistov, L. A. Prozorova, A. Prokofiev, and  W. A\ss mus,
Phys. Rev. B \textbf{76}, 014440 (2007).
\bibitem{neutron1} T. Masuda, M. Hagihala, Y. Kondoh, K. Kaneko, and N. Metoki,
J. Phys. Soc. Jpn. \textbf{80}, 113705 (2011).
\bibitem{neutron2} M. Mourigal, M. Enderle, B. F\aa k, R. K. Kremer, J. M. Law, A. Schneidewind, A. Hiess, and A. Prokofiev,
Phys. Rev. Lett. \textbf{109}, 027203 (2012).
\bibitem{NMR2} K. Nawa, M. Takigawa, M. Yoshida, and K. Yoshimura, J. Phys. Soc. Jpn. \textbf{82}, 094709 (2013).
\bibitem{magnetization} L. E. Svistov, T. Fujita, H. Yamaguchi, S. Kimura, K. Omura, A. Prokofiev, A. I. Smirnov, Z. Honda, and M. Hagiwara, 
JETP Lett. \textbf{93}, 21 (2011).
\bibitem{HFNMR} 
N. B\"{u}ttgen, K. Nawa, T. Fujita, M. Hagiwara, P. Kuhns, A. Prokofiev, A. P. Reyes, L. E. Svistov, K. Yoshimura, and M. Takigawa, submitted to Phys. Rev. B.
\bibitem{NaCuMoO4OH} A. Moini, R. Peascoe, P. R. Rudolf, and A. Clearfield, Inorg. Chem. \textbf{25}, 3782 (1986).
\bibitem{PbZnVO4OH} M. M. Qurashi and W. H. Barnes, Am. Mineral. \textbf{39}, 416 (1954).
\bibitem{Highfieldmagnetization} K. Kindo, S. Takeyama,
M. Tokunaga, Y. H. Matsuda, E. Kojima, A. Matsuo, K. Kawaguchi, and H. Sawabe,
J. Low Temp. Phys. \textbf{159}, 381 (2010).
\bibitem{HFEPR} N. Nakagawa, T. Yamada, K. Akioka, S. Okubo, S. Kimura, and H. Ohta, Int. J. Infrared Millimeter Waves \textbf{19}, 167 (1998). 
\bibitem{Johnston} D. C. Johnston, R. K. Kremer, M. Troyer, X. Wang, A. Kl\"{u}mper,
S. L. Bud\`ko, A. F. Panchula, and P. C. Canfield, Phys. Rev. B \textbf{61}, 9558 (2000).
\bibitem{ueda} A. N. Vasil'ev, L. A. Ponomarenko, H. Manaka, I. Yamada, M. Isobe,and Y. Ueda, Phys. Rev. B \textbf{64}, 024419 (2001). 
\bibitem{HTSE} A. B\"{u}hler, N. Elstner, and G. S. Uhrig, Eur. Phys. J. B \textbf{16}, 475 (2000).
\bibitem{ALPS} B. Bauer, L. D. Carr, H. G. Evertz, A. Feiguin, J. Freire, S. Fuchs, L. Gamper, J. Gukelberger, E. Gull, S. Gurtler, A. Hehn, R. Igarashi, S. V. Isakov,
D. Koop, P. N. Ma, P. Mates, H. Matsuo, O. Parcollet, G. Pawlowski, J. D. Picon, L. Pollet, E. Santos, V. W. Scarola, U. Schollw\"{o}ck, C. Silva, B. Surer, S. Todo,
S. Trebst, M. Troyer, M. L. Wall, P. Werner, and S. Wessel, J. Stat. Mech. P05001 (2011);
A. F. Albuquerque, F. Alet, P. Corboz, P. Dayal, A. Feiguin, S. Fuchs, L. Gamper, E. Gull, S. G\"{u}rtler, A. Honecker, R. Igarashi, M. K\"{o}rner,
A. Kozhevnikov, A. L\"{a}uchli, S. R. Manmana, M. Matsumoto, I. P. McCulloch, F. Michel, R. M. Noack, G. Pawlowski, L. Pollet, T. Pruschke, U. Schollw\"{o}ck,
S. Todo, S. Trebst, M. Troyer, P. Werner, and S. Wessel, J. Magn. Magn. Mater. \textbf{310}, 1187 (2007).
\bibitem{KG} V. H. Crawford, H. W. Richardson, J. R. Wasson, D. J.  Hodgson, and W. E. Hatfield, Inorg. Chem. \textbf{15}, 2107 (1976).
\bibitem{ang} Y. Mizuno, T. Tohyama, S. Maekawa, T. Osafune, N. Motoyama, H. Eisaki, and S. Uchida, Phys. Rev. B \textbf{57}, 5326 (1998).
\end{thebibliography}
\end{document}